\documentclass[12pt]{article}
\usepackage[margin=2cm]{geometry}
\usepackage{comment}
\usepackage{amsmath,amssymb,extarrows,mathtools,graphicx,subfigure,setspace}
\usepackage{cite}
\usepackage{slashed}
\usepackage{color}
\makeatother

\newcommand{\be}{\begin{equation}}
\newcommand{\bea}{\begin{eqnarray}}
\newcommand{\eea}{\end{eqnarray}}
\newcommand{\ba}{\begin{array}}
\newcommand{\ea}{\end{array}}
\newcommand{\ee}{\end{equation}}
\newcommand{\bes}{\begin{equation*}}
\newcommand{\beas}{\begin{eqnarray*}}
\newcommand{\eeas}{\end{eqnarray*}}
\newcommand{\bas}{\begin{array*}}
\newcommand{\eas}{\end{array*}}
\newcommand{\ees}{\end{equation*}}

\numberwithin{equation}{section}
\begin{document}
\onehalfspacing
\noindent
\begin{titlepage}
\hfill
\vbox{
    \halign{#\hfil         \cr
           IPM/P-2013/nnn \cr
                      } % end of \halign
      }  % end of \vbox
\vspace*{20mm}
\begin{center}
{\Large {\bf Entanglement Thermodynamics}\\
}

\vspace*{15mm} \vspace*{1mm} {Mohsen Alishahiha$^a$,\;  Davood
Allahbakhshi$^b$\; and\; Ali Naseh$^b$ }

 \vspace*{1cm}
{\it $^a$School of physics, \\ $^b$ School of Particles and Accelerators,\\
%P.O. Box 19395-5531, Tehran, Iran \\ }
%$^b${\it
 Institute for Research in Fundamental Sciences (IPM)\\
P.O. Box 19395-5531, Tehran, Iran \\ }

\vspace*{.4cm}

{E-mails: {\tt alishah@ipm.ir, allahbakhshi@ipm.ir, naseh@ipm.ir}}%

\vspace*{2cm}
\end{center}
\begin{abstract}
We study entanglement entropy for an excited state by making use of
the proposed holographic description of the entanglement entropy.
For a sufficiently small entangling region and with reasonable
identifications we find an equation between entanglement entropy and
energy which is reminiscent of the first law of  thermodynamics. We
then suggest four statements which might be thought of as four laws
of entanglement thermodynamics.

\end{abstract}
\end{titlepage}
%%%%%%%%%%%%%%%%%%%%%%%%%%%%%%%%%%%%%%%%%%%%%%%%%%%%%%%%%%%%%%%%%%%%%%%%%%%%%%
%%%%%%%%%%%%%%%%%%%%%%%%%%%%%%%%%%%%%%%%%%%%%%%%%%%%%%%%%%%%%%%%%%%%%%%%%%%%%%
\section{Introduction}
Thermodynamics provides a  useful tool to study a system when it is
in thermal equilibrium. In this limit the physics may be
described in terms of few macroscopic quantities such as energy,
temperature, pressure,  entropy and certain chemical potential  if
the system is charged. There  are also laws of  thermodynamics which
describe how these quantities behave under various conditions. In
particular the first law of thermodynamics which is a version of the
law of the conservation of energy, tells us how the entropy changes
as one changes the energy of the system.

We note, however, that there are several interesting phenomena which
occur when the system is far from thermal equilibrium. In fact a
rapid change in a system, such as quantum quenches, may  bring the
system out of equilibrium and indeed it is interesting to study the
thermalization process of this quantum system.

Although when the system is far from thermal equilibrium the
thermodynamical quantities may not be well defined, it is still
possible to compute the entanglement entropy. Therefore entanglement
entropy may provide a useful quantity to study excited quantum
systems which are far from thermal equilibrium.  Of course for a
generic quantum system it is difficult to compute the entanglement
entropy. Nevertheless, at least, for those quantum systems which
have holographic descriptions, one may use the holographic
entanglement entropy\cite{RT:2006PRL} to explore the behavior of the
system.

Another quantity which can be always  defined is the energy (or
energy density) of the system. It is then natural to pose the
question whether there is a relation between the entanglement
entropy of an excited state and its energy. Such a question has
recently been addressed in a certain situation using holographic
entanglement entropy in \cite{Bhattacharya:2012mi} where it was
shown that for a sufficiently small subsystem, the change of the
entanglement entropy is proportional to the change of the energy of
the subsystem. The proportionality constant is indeed given by the
size of the entangling region. To make contact with the first law of
thermodynamics the {\it entanglement temperature } has been
identified with the inverse of the  size of the entangling
region\cite{Bhattacharya:2012mi}.

The aim of the present article is to further explore a possible
generalization of the laws of thermodynamics for quantum
entanglement (see also \cite{Fursaev:2010ix}). More precisely using the holographic description of
the entanglement entropy at a certain limit for a specific model we
suggest several statements which are  reminiscent of laws of
thermodynamics. This may be thought of as {\it entanglement
thermodynamics}. We must admit that our results are based on an
explicit example and therefore one should be cautious to consider
them as a general framework.

The paper is organized as follows. In the next section using the
holographic description of the entanglement entropy we will derive
an equation which might be considered as the first law of
entanglement thermodynamics which make a connection between
entanglement entropy, energy and entanglement pressure (to be
defined later). In section three we investigate the universal
feature of the first law. In section four we suggest several
statements which could be considered as other laws of the
entanglement thermodynamics. The last section is devoted to
discussions.

{\bf Note}: While we were preparing to submit our paper we were aware of the paper
\cite{paper} where  the entanglement pressure has also been studied.
Moreover  after submitting of our paper to arXiv another
paper\cite{Blanco:2013joa}
 appeared where the relative entropy has been studied. In this
 context the contribution of the pressure to the change of the entanglement
entropy for an excited state has also been discussed.

%%%%%%%%%%%%%%%%%%%%%%%%%%%%%%%%%%%%%%%%%%%%%%%%%%%%%%%%%%%%%%%%%%%%%%%%%%%%%%
\section{First law of the entanglement thermodynamics}

According to the AdS/CFT correspondence\cite{M:1997} gravity on an
asymptotically locally AdS  provides a holographic description for a
strongly coupled quantum field with a UV fixed point. In this
context the information of quantum state in the dual field theory is
encoded in the bulk geometry. In particular the AdS geometry is dual
to the ground state of the dual conformal field theory.

Exciting the dual conformal field theory from the ground state to an
excited state holographically corresponds to  modifying the bulk
geometry from an AdS solution to a general asymptotically local AdS
solution. For example if one excites the ground state by heating up
the system, the bulk gravity would promote to an AdS black hole.

The aim of this section is to compute the entanglement entropy of an
excited state\footnote{Entanglement entropy for excited states in two dimensions has also been
studied in \cite{{Mosaffa:2012mz},{Astaneh:2013gp}}.} for the case where the entangling region is
sufficiently small (below we make it precise what we mean by
sufficiency small). Since the entanglement entropy for a small
subsystem would probe the UV region of the theory, from an  holographic
point of view one only needs to know the asymptotic behavior of the
bulk geometry.

On the other hand it is known that  the most general form of the
asymptotically locally AdS may be written in terms of the
Fefferman-Graham coordinates as follows \be\label{metric}
ds_{d+1}^2=\frac{R^2}{r^2}\bigg(dr^2+g_{\mu\nu} dx^\mu dx^\nu\bigg),
\ee where $g_{\mu\nu}=\eta_{\mu\nu}+h_{\mu\nu}(x,r)$ with \be
h_{\mu\nu}(x,r)=h^{(0)}_{\mu\nu}(x)+h^{(2)}_{\mu\nu}(x)r^2+\cdots+r^d\left(h^{(d)}_{\mu\nu}(x)
+{\hat h}^{(d)}_{\mu\nu}(x) \log r\right)+\cdots \ee The log term is
present for even $d$. The information about the excited state (or
the bulk geometry) is encoded in the function $h_{\mu\nu}(x,r)$.
This deformation of pure AdS geometry could be caused by heating up
the background or by the back reaction of other fields in the
model\footnote{Here we will only consider Einstein gravity. For
other gravitational models such as those with higher derivative
terms we have other powers in the  asymptotic behavior of the
metric.} .
 Of course in what follows we do not need the
explicit form of this function.

 To proceed let us fix our notation by reviewing the  computations of the holographic
entanglement entropy for a strip in an AdS geometry. A  $d+1$
dimensional  AdS solution in the Poincar\'e coordinates may be
written  as follows
\be ds^2=\frac{R^2}{r^2}(dr^2+\eta_{\mu\nu}
dx^\mu dx^\nu),\;\;\;\;\;\;\mu,\nu=0,1,\cdots, d-1.
 \ee
 Let us
consider an entangling region  in the shape of  a  strip with the
width of $\ell$ given by
 \be\label{strip} -\frac{\ell}{2}\leq
x_1\leq \frac{\ell}{2},\;\;\;\;\;\;\; 0\leq x_i\leq
L,\;\;\;\;\;i=2,\cdots, d-1. \ee
Following \cite{RT:2006PRL} the
holographic entanglement entropy may be computed by minimizing a
codimension two hypersurface in the bulk geometry which ends on the boundary of
 the above strip. Then the
entanglement entropy is the minimal surface divided by $4G_N$ where
$G_N$ is the Newton's constant of the bulk gravity.

In the present case where the background is an AdS$_{d+1}$ geometry, assuming the bulk extension
of the surface to be parameterized by $x_1=x(r)$, the corresponding area is given by
\be
A_0=R^{d-1}L^{d-2}\int dr \frac{\sqrt{1+x'^2}}{r^{d-1}}.
\ee
By making use of the standard procedure one may minimize the area to get\cite{RT:2006PRL}
\be\label{AdS0}
\ell=2\int_0^{\tilde{r}_t}dr \frac{(r/\tilde{r}_t)^{d-1}}{\sqrt{1-(r/{\tilde r}_t)^{2(d-1)}}},\;\;\;\;\;\;\;\;S^{(0)}_E(\tilde{r}_t)=2\frac{R^{d-1}L^{d-2}}{4G_N}
\int_\epsilon^{{\tilde r}_t}\frac{dr}{r^{d-1}\sqrt{1-(r/\tilde{r}_t)^{2(d-1)}}},
\ee
where $\tilde{r}_t$ is turning point and $\epsilon$ is a UV cut off. Thus one gets
\bea\label{GEAdS}
S^{(0)}_E=\frac{ L^{d-2} R^{d-1}}{2(d-2)G_{N}}\bigg{[}\frac{1}{\epsilon^{d-2}}-2^{d-2}\pi^{(d-1)/2}
\left(\frac{\Gamma\left(\frac{d}{2d-2}\right)}{\Gamma\left(\frac{1}{2d-2}\right)}\right)^{d-1}
\frac{1}{\ell^{d-2}}\bigg{]},
\eea

Now let us consider a deformation of  the AdS geometry which in turn
corresponds to dealing with an excited state in the dual field
theory. The aim is to compute the entanglement entropy of the strip
\eqref{strip} for an excited state when the width of strip is
sufficiently small so that  only the UV regime of the system will be
probed. Holographically this means that one needs to compute a
codimension two hypersurface on the asymptotically locally AdS
geometry which is given by the equation \eqref{metric} in the
Fefferman-Graham coordinates.

It is important to note that the deviation from AdS geometry in the
bulk does not need to be small. Indeed in what follows, using the
notation of the Fefferman-Graham coordinates, we assume that
$h_{\mu\nu}^{(n)} \ell^n\ll 1$\footnote{For a thermal geometry it corresponds to the condition
of $T\ell\ll 1$ where $T$ is the temperature. See \cite{Fischler:2012ca} for similar computations.}.
In fact this is what we mean by
``sufficiently small''. Note that in this limit, practically one
needs to compute the minimal surface up to order of ${\cal O}(h)$.

For the above strip the induced metric in the Fefferman-Graham
coordinates is \bea
ds^2=\frac{R^2}{r^2}\bigg((1+g_{11}x'^2)dr^2+2g_{1i}x' dr dx^i+
g_{ij}dx^idx^j\bigg). \eea Therefore to find the holographic
entanglement entropy one needs to minimize the following area \be
A=R^{d-1}\int d^{d-2}x dr\;
\frac{\sqrt{g(r)\;(1+G(r)\;x'^2)}}{r^{d-1}} \ee where
$g(r)=\det(g_{ij})$ and $G(r)=g_{11}-g_{1i}g^{-1}_{ij}g_{j1}$.

 To proceed we consider the case where the solution is static. Moreover to find analytic expressions
for our results we will assume that the components of the asymptotic
metric are independent of $x_1$, the direction the width of strip is
extended\footnote{Actually as far as the leading order behavior is
concerned, which is indeed the case for sufficiently small
entangling region, both assumptions may be dropped.}. With these
assumptions the equation of motion of $x$ leads to a constant of
motion \be\label{eq0}
\bigg(\frac{R}{r}\bigg)^{d-1}\frac{\sqrt{g(r)}\;G(r)\;x'}{\sqrt{1+G(r)\;x'^2}}={\rm
const}=c, \ee so that \be x'=\frac{c}{\sqrt{G(r)\bigg[g(r)\;G(r)\;
(\frac{R}{r})^{2(d-1)}-\;c^2 \bigg]}}. \ee The constant $c$ may be
found in terms of  the value of the left hand side of the equation
\eqref{eq0} evaluated at
 a specific point. Usually the specific point is chosen to be the turning point where $x'$ diverges.
Denoting  by $r_t$ the turning point,  one finds
\be
c^2=g(r_t)\;G(r_t)\;\left(\frac{R}{r_t}\right)^{2(d-1)} \ee
It is
then straightforward to find the  entanglement entropy and the width
of the strip as follows
 \bea\label{SL}
 &&S_{E}=\frac{1}{2G_N}\int
_0^{r_t}d^{d-2}x dr\; \big( \frac{R}{r} \big)^{2(d-1)}\;\sqrt{
\frac{g(r)^2 G(r)}{g(r)\;G(r)\; (\frac{R}{r})^{2(d-1)}-\;c^2}}\cr
&&\cr &&\ell=2\int _0^{r_t}dr\;\frac{c}{\sqrt{G(r)\big[g(r)\;G(r)\;
(\frac{R}{r})^{2(d-1)}-\;c^2\big]}} \eea
To evaluate  the above
expressions we note that at leading order one has \be g(r)=1+{\rm
Tr}(h_{ab})-h_{11}+{\cal O}(h^2),\;\;\;\;\;\;\;\;G(r)=1+h_{11}+{\cal
O}(h^2), \ee where $a,b=1,2,\cdots,d-1$. So that $g(r)G(r)=1+{\rm
Tr}(h_{ab})+{\cal O}(h^2)$.

In what follows in order to simplify the expressions, it is found useful to define the following
parameters
\be
\gamma(r)={\rm Tr}(h_{ab}),\;\;\;\;
\beta(r)=h_{11},\;\;\;\;f(r,r_t)=\sqrt{1-\left(\frac{r}{r_t}\right)^{2(d-1)}}.
\ee
In this notation  at the first order in $h$ one arrives at
\be
\ell=\int_0^{r_t}\frac{(r/r_t)^{d-1}}{f(r,r_t)}\bigg[
2+\frac{\gamma(r_t)-\gamma(r)}{f^2(r,r_t)}
-\beta (r)
 \bigg]dr
 \ee
It is worth noting that the aim was to compute the entanglement
entropy for an excited and compare it with the ground state which is
represented by an AdS solution. Actually we are interested in the
change  of the entanglement entropy caused by the change of the
state. Therefore we keep the  entangling surface fixed. Since $\ell$
is kept fixed while the geometry is deformed the turning point
should also be changed. Indeed assuming $r_t={\tilde r}_t+\delta
r_t$ with ${\tilde r}_t$ being the turning point for the pure AdS
case, one finds \be
 \delta
 r_t=-\frac{1}{2a_d}\int_0^{\tilde{r}_t}\frac{(r/\tilde{r}_t)^{d-1}}{f(r,\tilde{r}_t)}\bigg[\frac{\gamma(\tilde{r}_t)-\gamma(r)}{f^2(r,\tilde{r}_t)}-\beta
 (r)\bigg]dr
\ee
 where
\be a_d=\int_0^1\frac{\xi^{d-1}}{\sqrt{1-\xi^{2(d-1)}}}d\xi, \ee
Moreover the width of the strip $\ell$ is the same as that in pure
AdS geometry that is given by the equation \eqref{AdS0} which is
$\ell=2{\tilde r}_t a_d$. $\tilde{r}_t$ is the turning point for
pure AdS geometry.

It is now straightforward to compute the entanglement entropy up to order of ${\cal O}(h)$.
In fact expanding the expression of the entanglement entropy one finds\footnote{Note that
we could have found this result by replacing $r_t$ with $\tilde{r}_t$ in the first equation of
\eqref{SL},  then expanding at the first order in $h$ and requiring  $\delta r_t=0$. We would like to
thank
the referee for his/her comment on this point.}
\bea
S_{E}=S^{(0)}_E({\tilde r}_t)+\frac{R^{d-1}}{4 G_N}\int_0^{\tilde{r}_t} dr\;d^{d-2}x\;\frac{\gamma(r)-f^2(r,\tilde{r}_t)\beta(r)}{r^{d-1}f(r,\tilde{r}_t)},
\eea
 where $S^{(0)}_E({\tilde r}_t)$ is the holographic entanglement entropy for the strip in a pure AdS$_{d+1}$ geometry given in the equation \eqref{AdS0}.

By making use of the Fefferman-Graham expansion for the asymptotic
form of the metric one arrives at
\bea
\Delta S_{E}=\frac{R^{d-1}}{4
G_N}\int_0^{\tilde{r}_t} dr\left(\Gamma^{(0)}+\Gamma^{(2)}r^2+
\cdots+ \Gamma^{(d)}r^d+{\hat \Gamma}^{(d)}r^d\ln r\right), \eea
where the change of the entanglement entropy is defined by $\Delta
S_E=S_E-S^{(0)}_E({\tilde r}_t)$, and also
\bea
&&\Gamma^{(n)}=\frac{\int d^{d-2}x\; {\rm
Tr}(h^{(n)}_{ab})}{r^{d-1}f(r,\tilde{r}_t)}-\frac{f(r,\tilde{r}_t)}{r^{d-1}}
\int d^{d-2}x\; h^{(n)}_{11}\cr &&\tilde{\Gamma}^{(d)}=\frac{\int
d^{d-2}x\; {\rm Tr}({\hat
h}^{(d)}_{ab})}{r^{d-1}f(r,\tilde{r}_t)}-\frac{f(r,\tilde{r}_t)}{r^{d-1}}
\int d^{d-2}x\; {\hat h}^{(d)}_{11}.
\eea
 Using this expansion it
is straightforward to perform the integration over $r$. Indeed for
$d>2$ one finds
\bea\label{eq1} \int_\epsilon^{\tilde{r}_t}
dr\;\Gamma^{(n)}r^n&=&\frac{1}{(d-2-n)\epsilon^{d-2-n}} \int
d^{d-2}x\bigg ({\rm Tr}(h^{(n)}_{ab})-h^{(n)}_{11}\bigg)\cr&&\cr
&&-\frac{F(d-1,d-1-n)}{\tilde{r}_t^{d-2-n}} \int d^{d-2}x \bigg({\rm
Tr}(h^{(n)}_{ab})-\frac{d-1}{n+1}h^{(n)}_{11}\bigg)\cr &&\equiv
\frac{1}{(d-2-n)\epsilon^{d-2-n}}\;N^{(n)}
+\frac{1}{\tilde{r}_t^{d-2-n}}\;M^{(n)} , \eea
where $\epsilon$ is a
UV cut off, and
\be
F(m,n)=\frac{{}_2F_1\left(\frac{1}{2},\frac{1-n}{2m},\frac{2m+1-n}{2m},1\right)}{n-1},
\ee
 with $_2F_1$ being the hypergeometric function. Note that for
even $d$ and $n=d-2$ one finds just a logarithmic divergence as
$N^{(d-2)}\;\ln\frac{\epsilon}{\tilde{r}_t}$ while for odd $d$ and
$n=d-1$ the result is finite and is given by $M^{(d-1)}\;\tilde{r}_t$.
On the other hand for arbitrary $d$ for $n=d$ it leads to a finite
term  given by $\tilde{r}_t^2M^{(d)}$.
More precisely, using the fact
that in general  at leading order ${\rm Tr}(h^{(d)}_{\mu\nu})={\cal A}$ with ${\cal
A}$ being the trace anomaly one finds \be \int_0^{\tilde{r}_t}
dr\;\Gamma^{(d)}r^{d} =-F(d-1,-1)\;\tilde{r}_t^2 \int d^{d-2}x
\bigg(h^{(d)}_{tt}+{\cal A}-\frac{d-1}{d+1}h^{(d)}_{11}\bigg). \ee
 Note that for odd $d$ the anomaly term is zero. One should add that when $d$ is an even number we have another term coming from
${\hat \Gamma}^{(d)}$ which can similarly be calculated leading to
an $\ln\tilde{r}_t$ contribution to the entanglement entropy.

Having found these expressions  and taking into account that
$\ell=2\tilde{r}_t a_d$, one can find the variation of the
entanglement entropy, $\Delta S_{E}$, as a function of $\ell$. More
precisely for odd $d$ one has
 \bea \Delta
S_{E}&=&\frac{R^{d-1}}{4G_N}\sum_{n<d-2}\left(\frac{1}{(d-2-n)\epsilon^{d-2-n}}\;N^{(n)}
+\frac{(2a_d)^{(d-2-n)}}{\ell^{d-2-n}}\;M^{(n)} \right)
+\frac{R^{d-1}M^{(d-1)}}{8 G_N a_d}\;\ell\cr &&\cr
&&-\frac{R^{d-1}F(d-1,-1)}{16 a_d^2G_N}\;\ell^2\;\int
d^{d-2}x\;\bigg(h^{(d)}_{tt}-\frac{d-1}{d+1}h^{(d)}_{11}\bigg)+\cdots,
\eea
 while for even $d$ one gets \bea \Delta
S_{E}&=&\frac{R^{d-1}}{4G_N}\sum_{n<d-2}\left(\frac{1}{(d-2-n)\epsilon^{d-2-n}}\;N^{(n)}
+\frac{(2a_d)^{d-2-n}}{\ell^{d-2-n}}\;M^{(n)} \right)
+\frac{R^{d-1}N^{(d-2)}}{4G_N}\;\ln\frac{2\epsilon a_d}{\ell}\cr
&&\cr &&-\frac{R^{d-1}F(d-1,-1)}{16 a_d^2G_N}\;\ell^2\;\int
d^{d-2}x\;\bigg(h^{(d)}_{tt}+{\cal
A}-\frac{d-1}{d+1}h^{(d)}_{11}\bigg)\cr
&&-\frac{R^{d-1}F(d-1,-1)}{16 a_d^2G_N}\;\ell^2\ln\frac{\ell}{2a_d}\;\int
d^{d-2}x\;
\bigg(\hat{h}_{tt}^{(d)}-\frac{d-1}{d+1}\hat{h}^{(d)}_{11}\bigg)+\cdots.
\eea Here we have used the fact that ${\rm
Tr}(\hat{h}_{\mu\nu}^{(d)})=0$.

 Entanglement entropy is divergent due to short range interactions near the
boundary of the entangling surface and thus a UV cut off is needed.
The coefficient of the most divergent term is proportional to the
area of the entangling surface. Although in  the  expressions of the
change of the entanglement entropy, $\Delta S_{E}$, the divergent
term coming from the AdS geometry has already been subtracted, it
still has divergent terms whose coefficients are given by $N^{(n)}$
for $n\leq d-2$. It is worth to note that all of these terms are
given in terms of $h_{\mu\nu}^{(0)}$ and its derivatives whose
precise form may be found from the holographic renormalization
procedure\cite{deHaro:2000xn}. Therefore as soon as the boundary
becomes curved we have extra divergent terms due to the extrinsic
curvature of the boundary. Note that for the flat boundary all
$N^{(n)}$'s  vanish.

On the other hand the most relevant terms of the finite parts of
$\Delta S_{E}$  are given in terms of $M^{(n)}$ which is again given
in terms of $h_{\mu\nu}^{(0)}$ and its derivatives which vanish for
the flat boundary.  There is also one extra important term which
comes  from the $d$ th order of the Fefferman-Graham expansion. Note
that this term  does not depend on $h^{(0)}_{\mu\nu}$ and therefore
its contribution remains non-zero even for the flat boundary. Indeed
it has a very interesting feature as we explore below.

When one excites the ground state to an excited state, the energy of
the system is  increased and generally one gets non-zero expectation
value for the energy momentum tensor. By making use of the
holographic renormalization one can compute this expectation value.
Indeed  one has \cite{deHaro:2000xn} \be \langle T_{\mu\nu}\rangle
=\frac{dR^{d-1}}{16\pi G_N}\;h_{\mu\nu}^{(d)}. \ee In other words
from the dual gravity  point of view the expectation value of  the
energy momentum tensor is given by $h_{\mu\nu}^{(d)}$, which is
exactly the extra contribution to the holographic entanglement
entropy as we just mentioned. Therefore the extra non-trivial
contribution to the entanglement entropy is coming from expectation
value of the energy-momentum tensor which does depend on the excited
state we are considering. More precisely one finds

 \be\label{DS}
\Delta S_{E}^{\rm finite}=\sum_n(\cdots)\frac{M^{(n)}}{\ell^{d-2-n}}-\frac{\pi F(d-1,-1)\ell}{a_d^2 d}
\left(\Delta E-\frac{d-1}{d+1}
\int \Delta P_{x} dV_{d-1}+\frac{d R^{d-1}}{16\pi G_N}\int  {\cal A} dV_{d-1}\;\right)+\cdots,
\ee
where $(\cdots)$ stands for a numerical factor and  $dV_{d-1}=\ell d^{d-2}x$. Moreover the energy
 and {\it entanglement pressure} are defined by
\be \Delta E=\int dV_{d-1}  \langle T_{tt}\rangle
,\;\;\;\;\;\;\;\Delta P_{x}=\langle T_{11}\rangle. \ee It is worth
mentioning that  since in general the system is not in the thermal
equilibrium, the pressure $P_{x}$ should not be identified with that
in the thermodynamics and indeed it was the reason we called it {\it
entanglement pressure}. Note also that only entanglement pressure
normal to the entangling surface appears in the finite part of the
change of the entanglement entropy.

 For the case of $h^{(0)}_{\mu\nu}=0$  where the geometry is  asymptotically AdS solution one
has
\be
h_{\mu\nu}(x,r)=h^{(d)}_{\mu\nu}(x)\;r^d.
\ee
Note that in this case since the boundary is flat the anomaly term is zero and therefore the equation \eqref{DS} reads
\be
\Delta S_{E}=\frac{\pi \ell}{2d}\frac{C_1}{C_0^2}\left(\Delta E-\frac{d-1}{d+1} \int dV_{d-1} \Delta p_x\right),
\ee
where
\be
C_0=\sqrt{\pi}\;\frac{\Gamma\left(\frac{d}{2(d-1)}\right)}{\Gamma\left(\frac{1}{2(d-1)}\right)},\;\;\;\;\;\;\;
C_1=\sqrt{\pi}\;\frac{\Gamma\left(\frac{d}{d-1}\right)}{\Gamma\left(\frac{d+1}{2(d-1)}\right)},
\ee

Following \cite{Bhattacharya:2012mi} one may define {\it
entanglement temperature}  in terms of the width of the strip. More
generally the entanglement temperature is proportional to the inverse of
typical size of the entangling region and the proportionality
constant depends on the shape of the entangling region. In
particular in the  present case the corresponding temperature may be
given by $T_E=\frac{2dC_0^2}{\pi C_1}\;\frac{1}{\ell}$. Assuming
 $h^{(d)}_{\mu\nu}$ to be constant the above equation may be
recast to the following form\footnote{ It is important to note that
although for $d=2$ we find  logarithmic divergences, the final
result is the same.} \be \Delta E=T_{E}\Delta S_{E}+\frac{d-1}{d+1}
V_{d-1} \Delta p_x \ee where $V_{d-1}$ is the volume of the
entangling region. Due to its similarity with the first law of
thermodynamics we would like to consider this expression as {\it the
first law of entanglement thermodynamics}.

The way the energy and the entanglement pressure were defined
suggests that the holographic  equation of state should be given by
${\rm Tr}(h^{(d)}_{\mu\nu})={\cal A}$. In particular for the flat
boundary where ${\cal A}=0$ and when the solution is isotropic the
equation of state becomes $h^{(d)}_{tt}=(d-1)h^{(d)}_{11}$.  Using
the holographic renormalization and the definition of
 the entanglement pressure the equation of state in the dual field theory is ${\cal E}=(d-1) P_x$,
where ${\cal E}$ is energy density.  In this case the first law of
entanglement thermodynamics reads\footnote{In comparison with the
result of  \cite{Bhattacharya:2012mi} one has an extra
$\frac{d}{d-1}$ which is due to our definition of entanglement
temperature.}
\be T_E\Delta S_{E}=\frac{d}{d+1}\Delta E.
\ee
 An explicit example of such a situation is the AdS Schwarzschild background whose metric is given by
\be ds^2=\frac{R^2}{\rho^2}\left(-f(\rho)
dt^2+\frac{d\rho^2}{f(\rho)}+\sum_{i=1}^{d-1}dx_i^2\right),\;\;\;\;\;\;
f(\rho)=1-\left(\frac{\rho}{\rho_H}\right)^{d}.
\ee where $\rho_H$
is the radius of horizon. By making use of the coordinate
transformation
\be \frac{dr}{r}=\frac{d\rho}{\rho f^{1/2}}, \ee one
may recast the metric to the Fefferman-Graham coordinates as follows
\be
 ds^2=\frac{R^2}{r^2}(dr^2+g_{\mu\nu} dx^\mu dx^\nu),
\ee whose asymptotic behavior of the metric components are \be
g_{tt}=-1+h_{tt}^{(d)} r^d=-1+\frac{4(d-1)}{d}\rho_H^d
r^d,\;\;\;\;\;\;\; g_{aa}=1+h_{aa}^{(d)} r^d=1+\frac{4}{d}\rho_H^d
r^d. \ee Note that in this case one observes
$h^{(d)}_{tt}=(d-1)h^{(d)}_{aa}=\frac{4(d-1)}{d}\rho_H^d$. For an
entangling region given by the strip \eqref{strip} the energy and
the entanglement pressure are given by $\Delta
E=\frac{4(d-1)}{d}\rho_H^d V_{d-1}$ and $\Delta
P_x=\frac{4}{d}\rho_H^d$, respectively. Plugging these expressions
in the first law of the entanglement thermodynamics one can easily
find the change of the entanglement entropy for the AdS
Schwarzschild black hole as follows \be T_E\Delta
S_E=\frac{4(d-1)}{d+1}\rho_H^d V_{d-1}. \ee

%%%%%%%%%%%%%%%%%%%%%%%%%%%%%%%%%%%%%%%%%%%%%%%%%%%%%%%%%%%%%%%%%%%%%

\section{Universal features of the first law}

In the previous section in order to introduce the first law of the
entanglement thermodynamics we have considered the entanglement
entropy for a strip.  It is then natural to see to what extent the
resultant first law is universal. In this section we will consider
the holographic entanglement entropy for a system in the  form of a
sphere to address this question.

To proceed let us first write down the boundary metric at a fixed time in the
spherical coordinates \bea
ds^2=\frac{R^2}{r^2}(dr^2+g_{ij}dx^i
dx^j)=\frac{R^2}{r^2}(dr^2+g_{\rho\rho}d\rho^ 2+ 2\rho g_{\rho
\alpha}d\rho d\theta^\alpha + \rho^2 g_{\alpha\beta} d\theta^\alpha
d\theta^\beta ), \eea where \be g_{\rho\rho} =\Omega ^i
\;g_{ij}\;\Omega^j,\;\;\;\;\;\;\; \;\;\;g_{\rho\alpha}=\Omega^i
g_{ij} \frac{\partial\Omega^j}{\partial\theta^\alpha},\; \;
\;\;\;\;\;\;\;\;g_{\alpha
\beta}=\frac{\partial\Omega^i}{\partial\theta^\alpha}g_{ij}
\frac{\partial\Omega^j}{\partial\theta^\beta} \ee Here $\Omega^i$s
are the angular elements with the condition
$\sum_i\Omega^i\Omega^i=1$.

Now the aim is to study the entanglement entropy for a sphere with a
radius $\ell$ in the boundary.
%The extension of the region to
%the bulk will be parameterized by
Setting  $\rho = \rho(r)$  the induced
metric on the codimension two hypersurface in the bulk is given by
\be ds^2=\frac{R^2}{r^2}\bigg[(1+g_{\rho\rho}\rho
'^2)\;dr^2+2\rho\rho' g_{\rho\alpha}\;dr\;d\theta ^\alpha + \rho^2
g_{\alpha\beta}\;d\theta^\alpha d\theta^\beta\bigg] \ee
 Therefore to
compute the holographic entanglement entropy one needs to minimize
the following area \be A=R^{d-1}\int drd\Omega_{d-2}\;\rho
^{d-2}\;\frac{\sqrt{g\;(1+G\;\rho '^2)}}{r^{d-1}} \ee where
$g=\det(g_{\alpha\beta})$ and
$G=g_{\rho\rho}-g_{\rho\alpha}\;g^{-1}_{\alpha\beta}\;g_{\beta\rho}$.

Since in the present case the above expression treated as a one
dimensional action  does not have a constant of motion in order to
find $\rho$ one needs to solve its equation of motion \be
\bigg[\frac{1}{r^{d-1}}\frac{g G\rho^{d-2}\rho '}{\sqrt{g(1+G\rho
'^2)}}\bigg]'=(d-2)\rho ^{d-3} \frac{1}{r^{d-1}}\sqrt{g\;(1+G\rho
'^2)} \ee It is easy to check that for the ground state where the
dual gravity is given by an AdS$_{d+1}$ geometry a solution of the
above equation with the boundary condition $\rho_0(r=0)=\ell$ is $\rho_0 = \sqrt{\ell^2-r^2}$.
It is then evident that the turning point is also given by $\tilde{r}_t=\ell$. Note also that in
this case $G=1$ and \be
 g_{\alpha\beta}^{(0)}=\frac{\partial\Omega^i}{\partial\theta^\alpha}\delta_{ij} \frac{\partial\Omega^j}{\partial\theta^\beta}.
\ee
Following our previous study the aim is to find the entanglement entropy for an excited state for a sufficiently small entangling region.
 To do so, one needs to expand the expression for the area which at leading order it yields
\be
A(\rho , \ell)=A(\rho_0 , \ell)+\delta _g A(\rho_0 , \ell).
\ee
Here
\bea\label{change}
A(\rho_0 , \ell)=R^{d-1}\int\int_0^{\ell} dr  d\Omega_{d-2}\; \rho_0 ^{d-2}
\frac{\sqrt{g^{(0)}(1+{\rho'}_0^2) }}{r^{d-1}},
\eea
is the minimal area for the case where the dual theory is an AdS$_{d+1}$ geometry and
\bea
\delta_g A(\rho_0 , \ell)=\frac{R^{d-1}}{2}\int\int_0^{\ell} dr d\Omega_{d-2}\;
 \rho_0 ^{d-2}\frac{\sqrt{g^{(0)}(1+{\rho'}_0^2) }}{r^{d-1}}\bigg[ g_0^{\alpha\beta}\;h_{\alpha\beta}+\frac{\rho _0'^2}{1+\rho _0'^2}\;h_{\rho\rho} \bigg],
\eea
is the the first order correction to the minimal area due to  the deviation from the AdS geometry.  With these expressions it is easy to find the change of the
entanglement entropy as follows
\be \Delta
S_{E}=\frac{\delta_g A(\rho_0 , \ell)}{4G_N}=\frac{R^{d-1}}{8G_N}\int\int_0^{\ell}
\frac{(\ell^2-r^2)^{\frac{d-3}{2}}\ell}{r^{d-1}}
\bigg[{\rm
Tr}(h_{ab})-\frac{\ell^2-r^2}{\ell^2}h_{\rho\rho}
\bigg]\;dr\;d\Omega_{d-2}. \ee
%Note also that the radius of the
%entangling sphere is found to be $\ell=\tilde{r}_t$.

Now we need to use the Fefferman-Graham expansion for the metric to
find an expansion for the change of the entanglement entropy.
Actually the result has the same structure as that in the strip
case. Namely there are divergent terms which must be regulated by
introducing a UV cut off and they all vanish when the boundary is
flat.

Let us consider the case where $h_{\mu\nu}=h_{\mu\nu}^{(d)}r^d$ then
one arrives at \be T_E\;\Delta S_{E}=\Delta E-\frac{d-1}{d+1}\int
\Delta P_\rho\;dV_{d-1} \ee where $ T_E=\frac{d}{2\pi\ell}$ and
$dV_{d-1}=\rho^{d-2}d\rho\; d\Omega_{d-2}$. As we have already
mentioned in the case of the strip, it is important to note that
when the system is isotropic then there is a relation between
pressure and energy. In the present case where the entangling region
is an sphere this condition is automatically satisfied leading to
the relation of
% between
%the energy and pressure
%\be
$\Delta E=(d-1)\int \Delta P_\rho dV_{d-1}$.
%\ee
Therefore in this case the above equation reads \be
\tilde{T}_E\;\Delta S_{E}=\Delta E, \ee in agreement with
\cite{{Nozaki:2013vta},{Blanco:2013joa}}\footnote{We would like to
thank the referee for a comment on this point.}. Here
$\tilde{T}_E=\frac{d+1}{d} T_E$.

To conclude we note that although the numerical factor in the
definition of the entanglement temperature is different from that in
the strip case, the final form of the first law is the same. In
particular the numerical factor in front of the pressure term is
universal and only the entanglement pressure normal to the
entangling surface appears in the first law. Therefore the first law could be
 universal and the only shape dependent parameter is the numerical
factor in the definition of entanglement temperature.

%%%%%%%%%%%%%%%%%%%%%%%%%%%%%%%%%%%%%%%%%%%%%%%%%

\section{Laws of entanglement thermodynamics}

In the previous sections based on the  holographic description of
the entanglement entropy and for explicit examples we have found a
relation between entanglement entropy, energy and entanglement
pressure which using the similarity with the thermodynamics
could be thought of as the first law of entanglement thermodynamics.
It is then natural to pose the question whether there are other laws
similar to what we have  in the thermodynamics. The aim of this
section is to introduce other statements that could be identified as
other laws of
 entanglement thermodynamics.

There is a natural statement for the second law of entanglement
thermodynamics. It is indeed the strong subadditivity that  must be satisfied by the
entanglement entropy\cite{{Lieb:1973zz},{Lieb:1973cp}}.
According to the strong subadditivity for any given three  subsystems
$A$, $B$ and $C$ that do not intersect one has
\be S_{E(A\cup B)}+S_{E(C\cup B)}\geq S_{E(A\cup B\cup C)}+S_{E( B)}.
\ee
Note that by setting $B$ empty in the above expression, one arrives at
\be S_{E(A)}+S_{E(C)}\geq S_{E(A\cup C)}.
\ee

 It is worth noting  that although the entanglement entropy is
divergent due to UV effects, the  divergent parts of the
entanglement entropy drop from both sides. In fact this inequality
is also satisfied by the finite part of the entanglement entropy.
 It is known that holographic
entanglement entropy defined as a minimal surface in the bulk does
satisfy this inequality too\cite{Headrick:2007km}. Therefore one may suggest  this relation as the second
 law of the entanglement thermodynamics.
%It is worth mentioning that  this second
 %law of the entanglement thermodynamics is quite different in nature form that in the thermodynamics.
%While in the thermodynamics the second laws is closely related to the dynamics of the system,
%in the present case it has nothing to do with the dynamic.

So far our suggestions and statements  about the laws of
entanglement thermodynamics were based on rigorous computations. To
proceed for other possible laws we note that although we will use an explicit
example to explore them, we should admit that our suggestions are
based on speculation.

The most important part of our study is the definition of
the entanglement temperature. Although from dimensional analysis and
also from our experiences in thermodynamics and hydrodynamics it is
natural to consider the inverse of the typical size of the
entangling region as the temperature, having a non-universal numerical
factor in its definition makes us  wonder  to what extent it is a well
defined quantity.

 Of course as long as we are considering entangling regions with a fixed shape the numerical
factor is universal\cite{Bhattacharya:2012mi}. On the other hand having different shapes may
lead to a  puzzle as
to define the temperature. Apart from this ambiguity, in what follows for a fixed shape we suggest a statement
which could be considered as the zeroth law of entanglement thermodynamics.

Consider two entangling regions given by two strips with  widths
$\ell_1$ and $\ell_2$, respectively. If we bring these two regions
close together we get another strip whose width at most could be
$\ell_3= \ell_1+\ell_2$. Therefore we have the following inequality
for the entanglement temperatures before and after joining the
systems.
 \be \frac{1}{T_{1E}}+\frac{1}{T_{2E}}\geq \frac{1}{T_{3E}}.
\ee
 It is easy to argue that such a relation could also be
satisfied when the entangling regions are spheres.

Let us now proceed to introduce the third law of entanglement
thermodynamics. To do so, consider the finite part of the
entanglement entropy of a strip for an excited state. Using the
definition of the entanglement temperature, up to order of ${\cal
O}(T^{-2}_E)$, one gets

\be S_E^{\rm finite}=\frac{  R^{d-1}}{8G_{N}}\left[(\tilde{B}_0
L^{d-2}+B_0 M^{(0)})T^{d-2}_E +\sum_{n\geq1}B_n
M^{(n)}T^{d-2-n}_E\right]+\frac{1}{T_E}\left(\Delta
E-\frac{d-1}{d+1}\Delta P_x \Delta V_{d-1}\right) \ee
 where
$\tilde{B}_0, B_n$ are numerical factors. For sufficiently large
entanglement temperature (small size) the finite part of the
entanglement entropy diverges as\footnote{For $d=2$ it diverges
logarithmically.}
 \be
S_E^{\rm finite}\sim T^{d-2}_E. \ee Therefore in principle the
finite part of entanglement entropy goes to infinity for
sufficiently higher entanglement  temperature. We note, however,
that due to a natural UV cut off in the theory there is a natural
cut off for temperature preventing to get infinite entanglement
entropy.

Note that  as we increase the temperature, the  dominant divergent
parts comes from the
 ground state which
corresponds to the AdS geometry.  It is then possible to argue that
the above statement is also valid for other shapes of the entangling
region. We would like to suggest the above statement as the third
law of entanglement thermodynamics.

\section{Discussions}

In this paper based on the holographic description of the
entanglement entropy and within an explicit example we have
suggested four laws for the quantum entanglement entropy which are
reminiscent of the laws of thermodynamics. The corresponding laws of
entanglement thermodynamics may be summarized as follows.

\begin{itemize}

\item {\it Zeroth law}:  The entanglement temperature is proportional to the inverse of the typical size
of the entangling region and for two subsystems $A$ and $B$ one has
\be \frac{1}{T_{(A) E}}+\frac{1}{T_{(B) E}}\geq \frac{1}{T_{(A\cup
B) E}}. \ee

\item {\it First law}: There is a relation between the energy of the system and the entanglement entropy as
follows \be \Delta E=T_E\Delta S_E+\frac{d-1}{d+1} V_{d-1} \Delta
P_{\bot}, \ee where $\Delta P_{\bot}$ is the entanglement pressure
normal to the entangling surface.
\item {\it Second law}: Entanglement entropy enjoys strong subadditivity
\be
 S_{E(A\cup B)}+S_{E(C\cup B)}\geq S_{E(A\cup B\cup C)}+S_{E( B)}.
\ee

\item {\it Third law}: There is an upper bound on the entanglement temperature preventing to have an
infinite entanglement entropy.

\end{itemize}

In this paper we have considered entanglement entropy for a static case where the
corresponding background geometry was time independent. It is, however, possible to show
that the final results also hold for time dependent cases. Of course when we are dealing with
a time dependent geometry, in general,  one should use the
covariant holographic entanglement entropy in which the entanglement entropy is given by a codimension
two hypersurface which is extremal \cite{Hubeny:2007xt}.

We note, however, that as long as we are interested in a
sufficiently small subsystem we could still use the static solution
leading to the same result for the first law. The reason is as
follows. Consider a time dependent excited state above a vacuum
solution. From the bulk point of view it corresponds to a time
dependent deviation from an AdS solution. There are several sources
which contribute to the change of the holographic entanglement
entropy. The change may be caused by the change of the turning
point, the change of the solution and the change of the metric. The
interesting point is that at leading order which is what we are
interested in, the change of entanglement entropy is completely
given by the change of metric (see the equation \eqref{change}). In
other words one has\cite{Nozaki:2013vta} \be \Delta
S_E=\frac{1}{4G_N}\int d^{d-1}x \sqrt{\det(g^{(0)}_{\rm in})}
(g^{(0)}_{{\rm in}\; ab})^{-1}g^{(1)}_{{\rm in}\;ab}, \ee
 where
$g^{(0)}_{\rm in}$ and $g^{(1)}_{\rm in}$ are the induced metrics on
the codimension two hypersurface in the bulk for the cases of AdS
geometry  and the perturbation above it, respectively. It is, now,
clear that from the AdS case which is static one can read the first
law. Indeed the result is the same as that we considered in the
previous section. Therefore the first law we have introduced in this
paper may also be applied for the time dependent case (see for
example \cite{Nozaki:2013wia}). Of course to have the second law one
needs to further assume the null energy condition for the excited
state \cite{Callan:2012ip}.

Recently Lewkowycz and Maldacena \cite{Lewkowycz:2013nqa} have
introduced a generalized gravitational entropy for classical
Euclidean gravity solutions. More precisely consider metrics that
end on a boundary which has a direction with the topology of a
circle. Note that the solution is not  necessarily  symmetric under
the $U(1)$ rotation of the circle. Moreover the boundary need not to
be a true asymptotic boundary of the metric  and indeed it is just a
place where the boundary conditions are imposed. One may associate
an entropy to this solution. If the circle never shrinks in the
interior of the bulk geometry the corresponding entropy is zero. If
it does, the entropy is given  by the area of a codimension two
hypersurface in the bulk of the solution which,  for the Einstein
gravity, satisfies the minimal area condition.  In fact at this
hypersurface the circle  shrinks to zero size.

Using the results  of the present paper and the fact that when  solutions are symmetric under the
$U(1)$ rotation  the above construction reduces  to the Gibbons-Hawking computation of the black hole entropy,
one may wonder that there could be a {\it generalized laws of thermodynamics} for the
generalized gravitational entropy. It would be interesting to explore this possibility.

It is worth noting that besides the holographic entanglement entropy
there are other interesting quantities which have been  studied in
the literature. These quantities include the geometric
entropy\cite{{Fujita:2008zv},{Bah:2008cj}} and its generalization
when one has fractionalized charges \cite{Allahbakhshi:2013wk}. It
would also be interesting to see if the first law  can also be
obtained for these quantities.

%{\bf Note added:} After submitting of our paper to arXiv another
%paper\cite{Blanco:2013joa}
 %appeared where the relative entropy has been studied. In this
% context the contribution of the pressure to the change of the entanglement
%entropy for an excited state has also been discussed.

\section*{Acknowledgments}

We would like to thank A. Davody, D. Fursaev, M. R. Mohammadi
Mozaffar, A. Mollabashi, A. E. Mosaffa, T. Takayanagi and A. Vahedi
for useful comments and discussions.

\end{document}